\documentclass{ws-procs9x6}

\begin{document}

\title{Vladimir Naumovich Gribov: \\
Pieces of biography\footnote{\uppercase{B}ased on the
invited talk at the 4th \uppercase{V}.\uppercase{N}. \uppercase{G}ribov
\uppercase{M}emorial \uppercase{W}orkshop ``\uppercase{T}heoretical \uppercase{P}hysics
of \uppercase{XXI} \uppercase{C}entury'' (\uppercase{G}ribov-85),
17-20 \uppercase{J}une,  2015, \uppercase{C}hernogolovka, \uppercase{R}ussia.
\uppercase{T}o appear in the \uppercase{P}roceedings.}}


\author{Y\lowercase{a}.~I. AZIMOV }

\address{Petersburg Nuclear Physics Institute, \\National Research Center
``Kurchatov Institute'', \\
Gatchina, 188300, Russia \\
E-mail: azimov@thd.pnpi.spb.ru}

\maketitle

\abstracts{
The talk presents the main lines of biography of the prominent
physicist and bright personality. Also given is a necessarily brief
description  of Gribov's scientific work.}

This talk I would like to begin with some personal recollections.
I was lucky to work with V.N.~Gribov for 20 years. It was not
{\it under} Gribov, but just {\it with}. We have several common
papers, several of my papers were made according to Gribov's
suggestions. But most of my work were formally independent of Gribov.
Nevertheless, all my work of those 20 years, in more or less detail,
went through discussions with Gribov.

Of course, physics was not the only interest of Gribov, but,
undoubtedly, it was his main interest producing very strong emotions.
One could discuss with Gribov various physics problems, even rather
far from his own studies (a couple of most bright examples will
be mentioned below). But those discussions were not easy. Gribov was
hearing very attentively and was ready to immediately jump into battle
if something seemed incorrect to him. He had a rare feature: his
arguments were very hard to reject even when his position appeared unjust
after all (I think nobody can be just in all cases). The same
situation was both in private discussions and at seminars. That is
why talks at Gribov's seminar were very difficult for speakers. As
a result, some people feared to speak at his seminar. However, if
a person were sufficiently brave, the talk appeared very useful for
the speaker. After such trial the author himself began to better
understand his own work and results.

Gribov's name and his papers (at least some of them) are well known
today. But his biography is less known. Here I present its main
lines.

Vladimir (Russian nickname Volodya) Gribov was born on March 25, 1930 in Leningrad
(St.Petersburg, before 1914 and now). His farther died in 1938, at the time of the Great Terror
in the Soviet Union. ``Luckily'', his death was the result of a decease, and not because
of repression (in which case the whole family could be repressed as well). But the situation
was not simple: the mother stayed alone with two small children, Volodya and his younger sister.

Gribov's mother worked in one of Leningrad theaters (not as an actress). When Germany attacked
the Soviet Union in 1941, the family was evacuated from Leningrad together with the theater.
They were moving with the theater over Siberia, Far East, Urals. Nevertheless, Volodya was
continuing his school education, started in 1937, without any delay. Only in summer 1945
the family was able to return to Leningrad (to enter the city after the blockade, every person
needed to have a special permit). In 1947 Volodya finished here the school course. The natural
question arose: what to do after that?

Volodya was growing up in theatrical environment, and he dreamed to
become an actor, best of all, a cinema actor. However, when being in
senior school classes, he had a possibility to expose himself to
filming. As appeared, under camera he became ``frozen'' and lost his
natural mobility. After that, one of professional actors advised him
to choose some other speciality. At school, Volodya was quite
successful in physics and mathematics. He preferred the former.

In 1947 Gribov enrolled in Physical Faculty of the Leningrad
University. He was a student together with D.V.~Volkov (later, the
full member of the Ukrainian Academy of Sciences) and G.M.~Eliashberg (now the full member
of the Russian Academy of Sciences, the chief researcher of the Landau Institute
for Theoretical Physics). In 1950 Volkov was included into the special group
to study nuclear physics in more detail, and in 1951 was transferred to the Kharkov University.
Gribov also wished to be in that group, but was rejected.

In 1952 Gribov presented his diploma thesis considering interaction of two electrons
in Quantum Electrodynamics (under supervision of Yu.V.~Novozhilov). It was evaluated very high.
According to decision of the examination commission, Bulletin of the Leningrad University
published the paper~\cite{2e} based on this diploma work. It was the first publication
signed by V.N.~Gribov. Thus, in summer 1952 he was
graduated from the University with {\it diploma cum laude}.

The year 1952 in the Soviet Union was not favorable for persons of
jewish origin, one of which was Vladimir Naumovich Gribov. It was the last year for the trial
of the Jewish Antifashist Committee (after its end, more than 100 persons were sentenced,
more than 20 of them were shot). It was also the preparatory year for the ``case of physicians''
(``killers in white coats'') which was officially opened in January 1953. In such situation
the only job available for Gribov, a young physicist with {\it diploma cum laude}, appeared
the position of a physics teacher in an evening school, organized for working people who could not,
by any reason, complete their school education in the childhood. The salary was low, and year later
Gribov found additional part-time work.

Nevertheless, Volodya wished and was continuing to do science. He was able to contact with Professor
L.E.~Gurevich. To the beginning of 1954, they together prepared two papers, on properties of matter in
external fields, electric and gravitation. The papers were submitted to Journal of Experimental and
Theoretical Physics and accepted for publication~\cite{el,gr}. In addition, Gribov participated in the
theoretical seminar of the Leninigrad Physico-Technical Institute (now the A.F.~Ioffe Physico-Technical
Institute, further PTI) headed by I.M.~ Shmushkevich and K.A.~Ter-Martirosyan.

At last, in May 1954 (after Stalin's death and extinction of the ``case of physicians''), Gribov was able
to be employed by PTI as the senior laboratory assistant in the group for nuclear theory. It was headed
by I.M.~Shmushkevich, who was simultaneously an informal head of the Theoretical Department as a whole.
Now Gribov's progress was very fast. A year later he received the higher position of the junior
scientist. In March 1956 he presented his dissertation for the degree of Candidate of Sciences
(analog of PhD). It was concerned with interactions of neutrons with nuclei~\cite{dif,rot}. The problem
was suggested by K.A.~Ter-Martirosyan, but at the presentation of the work he emphasized that methods
for calculations were invented by Gribov himself.

After the defence of the dissertation, Shmushkevich and Ter-Martirosyan organized contacts of Gribov
with L.D.~Landau and I.Ya.~Pomeranchuk. Gribov began to go regularly to Moscow for participating
in Landau's seminar. Initially, Landau was sceptical in respect to Gribov (he said: ``I know one Gribov,
the theatrical actor, and this is enough''). But he rapidly changed his mind. In 1958, when the
Scientific Council of PTI discussed the next higher position for Gribov, the senior scientist,
it received the very favorable recommendation from Landau. Later, Gribov always considered Landau
as his main Teacher in theoretical physics.

In 1957, Shmushkevich invited Gribov to give lectures for students of the Leningrad Polytechnical Institute,
where he was the Chair of the Theoretical Physics Department. Later, Gribov began to lecture in his
{\it alma mater}, in the Leningrad University, and in 1968 he became the Professor of the University
(the highest scientific degree, Doctor of Sciences, necessary for this, Gribov received in 1964). In parallel,
Gribov presented lectures at various Physics Schools, both in the Soviet Union and (later) abroad.

Gribov's scientific work after receiving the Candidate degree became also very active and self-reliant.
More and more often he appeared a source of ideas for his colleagues. For instance, in 1958 he investigated
three-pion decays of the K-meson. Pair-energy distributions in the decay were shown to depend
on pion-pion scattering length~\cite{np-tau,jetp-tau}. For several years after that, the series of papers
were published by members of Shmushkevich's group (including Gribov himself) on various inelastic
near-threshold reactions. They could allow to obtain information on hadron interactions ({\it e.g.},
pion-pion ones) unreachable in conventional ways. Regretfully, corresponding experimental attempts,
as appeared, could not give definite results at those times because of insufficient quality
of experimental technique.

One of his further directions of interest (probably, under influence of Pomeranchuk) was high-energy
behavior of strong interactions. It was generally assumed for long time to be similar to classical
diffraction of light on black screen. Gribov showed that such behavior would be  inconsistent with
analytic properties of strong-interaction amplitudes~\cite{np-dif}. This work of 1961 initiated
international interest to his activity. Trying to overcome this diffraction difficulty, Gribov, partly
in collaboration with Pomeranchuk, developed reggeology, method of Regge poles and, later, cuts as well.
This direction was also actively supported by his colleagues in PTI. He became one of leaders of reggeology
not only in the Soviet Union, but in the whole world as well.

 At the same time, Gribov tried to use the known quantum field theories as test-grounds for studying
their high-energy properties. In Quantum Electrodynamics, as he showed, a special role play the
 so-called ``double-logarithmic terms''. That is why the group of enthusiasts of this approach (G.V.~Frolov,
V.G.~Gorshkov, and  L.N.~Lipatov first of all) was informally called ``the double-logarithmic academy''.
Their attempts were crowned by papers of Gribov and Lipatov~\cite{gl1,gl2} on summation of those
``double-logarithmic terms'', very famous now (they became a basis for studying the evolution of partons).

Gribov could efficiently discuss even such problems in which he was not previously active. For instance,
he did not work himself with weak interactions and, in particular, with neutrinos, but when encountered
with Pontecorvo's idea of neutrino oscillations Gribov immediately began to construct the corresponding
formalism. This resulted in the joint paper of V.N.~Gribov and B.M.~Pontecorvo~\cite{gr-pont}, considered now
to be classic in the neutrino physics.

There was one more, less known example. In his talk at one of Gribov's seminars, Ya.B.~Zeldovich explained
that a charged rotating black hole should lose energy by radiation, so its rotation should slow down.
When it stops, radiation, according Zeldovich, would stop as well.
Here Gribov interrupted him by statement that this is not correct,
radiation would continue. Zeldovich waved away this statement, and the seminar talk was continued without 
discussion of Gribov's suggestion (it seems, however, that their discussions on this problem went on later, 
but Zeldovich stayed rigid). At that seminar, I restored for myself the lines of Gribov's thought as follows. 
Gribov, as a physicist, had grown up mainly on quantum physics (in difference with Zeldovich), and he was 
aware quite well about the Schwinger effect: strong enough electric field, even homogeneous and static 
({\it i.e.}, large gradient of electromagnetic potential), generates electron-positron pairs due to quantum 
tunneling.  Near a black-hole horizon, there is very large gradient of the gravitational potential, which 
should analogously produce particle-antiparticle pairs.

Some time later, Zeldovich again talked at Gribov's seminar, now about the famous paper of Hawking. 
In particular, Zeldovich said: ``Volodya Gribov had tried to assure me that radiation of a black hole 
should continue, but I did not believe''. That is how he lost the interesting and important result.

Gribov ever tried to organize his whole knowledge into some consistent picture. As a rule, such approach 
is very useful. But not always. It may be not quite adequate if the picture needs strong changes. For Gribov, 
such was the case of quarks. The idea of quarks was publicized by M.~Gell-Mann and G.~Zweig in 1964 
on the base of resonance spectroscopy. It was further supported by the discovery of scaling in deep-inelastic 
scattering (in 1968) and, especially, by the unexpected discovery of $J/\psi$ (in 1974). However, Gribov did 
not believe yet in quarks and, correspondingly, was sceptical in respect to Quantum Chromodynamics (QCD), 
which arose in 1971--1972. It was not a question of taste; for his position Gribov had definite rational 
arguments. At that time, I asked him once on reasons of his disbelieving in quarks. He answered: ``The quarks 
should be strong-interacting objects; therefore, each of them should be surrounded by a pion cloud. Where is it?'' 
(I should confess, that a clear consistent answer to this question is still absent. Note also that such negative 
relation to quarks did not prevent Gribov from suggesting to E.M.~Levin and L.L.~Frankfurt in early days of quarks 
to compare meson and baryon cross sections in terms of quarks~\cite{lf}.)

However, in 1976, after discovery of charmed particles, predicted by the quark picture, Gribov changed his mind. 
He began to study quarks and QCD very intensively. And in 1977 he was able to find a new feature of QCD, known now 
as the Gribov horizon or Gribov copies~\cite{gr-YM1,gr-YM2}. It is interesting to note that many--many people 
worked with QCD to that moment, but existence of the horizon stayed unnoticed. Initially, Gribov hoped that it is 
just the horizon which determines the origin of confinement of quarks and gluons. But soon he came to conclusion 
of its insufficiency. Since then he worked hard trying various ways to understand and describe a still unknown 
mechanism of confinement.

The administrative career of Gribov may look successful. To 1962 the Gatchina site of PTI had the working nuclear 
reactor and the proton accelerator under construction. There appeared necessity to have there a separate Theoretical 
Department. Gribov was suggested to organize it. Later, in 1969, after the death of I.M.~Shmushktvich, Gribov was 
returned to the central part of PTI and became the head of its Theoretical Department. In 1971 the Gatchina site 
of PTI was transformed to be a new institute, Leningrad (now Petersburg) Nuclear Physics Institute (LNPI, now PNPI). 
All the activity on atomic nuclei and elementary particles was transferred from PTI to the new Institute. Its 
Theoretical Department was headed, of course, by Gribov. He was a rather good administrator, but, in my opinion, 
he disliked administrative duties which interfered with his scientific work. Meanwhile, those duties increased 
along with the Department becoming more populous. In 1980 Gribov moved from LNPI to L.D.~Landau Institute 
of Theoretical Physics in Chernogolovka (near Moscow). There were several reasons for this step, and one of them, 
as I think, was his wish to diminish necessary administrative duties.

As a physicist, V.N.~Gribov was recognized world-wide. He was a speaker at various conferences and schools 
in the Soviet Union and, sometimes, even abroad. Many foreign physicists, being in the Soviet Union, were eager 
to visit LNPI for discussions with Gribov. For his scientific work Gribov won various prizes, but he was most
strongly proud to be the first recipient of the L.D.~Landau Prize established in 1971 by the Academy of Sciences 
of the USSR, which honored his Teacher. In the same year, 1971, he became a member of the American Academy 
of Arts and Sciences. In the next year, 1972, he was elected (after several unsuccessful attempts) to be 
a corresponding member of the Academy of Sciences of the USSR. However, official position in respect to him 
was demonstrated by the fact that Gribov has never been elected to be a full member of the Soviet Academy.

After 1980, Gribov obtained the possibility to be part-time in Budapest, where his second wife, Julia Nyiri, 
worked in Physics Institute. This made somewhat easier his contacts with West physicists. And after 1990, 
longer abroad trips became possible, to be a visiting Professor of various Institutes and Universities, both
in Europe and in the US.

Active and intensive work of Gribov was unexpectedly interrupted by the acute stroke in 1997 during one 
of scientific conferences. He was taken to a hospital and, after stabilization of his state, was transported 
to another hospital, in Budapest. Even there he tried to continue investigations of confinement. The medical 
treatment looked successful, and the physicians planned that Gribov would be able soon to go home. However, 
on August 13, 1997 he passed away. His grave in Budapest is marked by the memorial that reminds a beautiful 
fading flower, with the simple epitaph
\begin{center}
{VLADIMIR GRIBOV\\
FIZIKUS$~$\\
1930--1997}
\end{center}
on the basement.

After Gribov's death, his works have not been forgotten. Just opposite, many of his papers are republished. 
His lectures. which were mainly written up in Russian, but not always published, are now collected, translated 
into English and published as books. Therefore, his results become available and well-known even for younger 
generations of physicists. If judging by references, the most famous and operative of those results seem to be 
the Gribov copies (they are especially essential now in lattice calculations) and DGLAP equations for evolution 
of partons (here G stays just for Gribov). There established are various stipends and prizes called by Gribov's 
name.

And yet there are Gribov's last papers or even notes. They are mainly not completed or, at least, not quite 
understood by the world community. Meanwhile, they tried various ways to solve the problem of confinement, 
one of the hottest problems in strong interactions. In some sense the situation may be similar to the fate 
of Einstein's last ideas. During his life, they looked to be out of mainstream, but now many of them feed 
new theoretical approaches. Such future is not excluded also for Gribov's last ideas. Look forward...

\section*{Acknowledgments}

This work is supported by the Russian Science Foundation (Grant No.14-22-00281).

\end{document}